\documentstyle[twoside,fleqn,espcrc2,epsf]{article}

\newcommand{\AmS}{{\protect\the\textfont2
  A\kern-.1667em\lower.5ex\hbox{M}\kern-.125emS}}

\title{Recent Lattice Results on the Light Quark Masses
}
\author{Paul B. Mackenzie
\address{Theoretical Physics Dept., Fermilab, P.O. Box 500, Batavia, IL  60510  USA (permanent address)}
\address{Center for Computational Physics, University of Tsukuba, Tsukuba,
Ibaraki 305, Japan}
}
       
\begin{document}

\begin{abstract}
 I discuss old and new determinations of the light quark masses
using lattice QCD.
Most lattice results using various approximations
 can be fit together in a simple picture which
is consistent with  lower values than conventionally supposed:
in the quenched approximation
for the strange quark mass in the $\overline{MS}$ scheme,
$\overline{m}_s(2\ {\rm GeV})	 = 	 95 (16)\ {\rm MeV}$, and
for the average of the $u$ and $d$ quark masses,
$\overline{m}_l(2\ {\rm GeV})	 = 	 3.6 (6)\ {\rm MeV.}$
An estimate of the effects of light quark loops yields answers lower still:
$\overline{m}_s({\rm 2\ GeV})$  in the range  54 - 92 MeV,
and $\overline{m}_l({\rm 2\ GeV})$ in the range 2.1 - 3.5 MeV.

\end{abstract}

\maketitle

\section{Introduction}
Among the most important applications of lattice QCD are the determinations
of the fundamental parameters of the standard model in the quark sector.
Of these, one of the most important is the overall scale of the light
quark masses.
It is one of the most poorly known of the parameters of the standard model
from prelattice methods.
The Particle Data Group estimates a range of a factor of three in allowed
values \cite{RPP96}:
\begin{eqnarray}
100\ {\rm MeV}&<&\overline{m}_s(1\ {\rm GeV}) <300\ {\rm MeV,}  \\
5\ {\rm MeV}&<&\overline{m}_d(1\ {\rm GeV}) <15\ {\rm MeV,\ and}	\\
2\ {\rm MeV}&<&\overline{m}_u(1\ {\rm GeV}) <8\ {\rm MeV.}
\end{eqnarray}
(I will use $\overline{m}_q$ to represent the running quark mass in 
the $\overline{MS}$ scheme.)
Global fits to standard model parameters are very sensitive to
such large variations.
It is also one for which lattice methods provide the only systematically
improvable determination.
This can be contrasted with the strong coupling constant $\alpha_s$, 
for example,
for which it is possible to imagine going to higher and higher energy
scattering experiments and extracting $\alpha_s$ with perturbation theory.
Lattice quark mass extractions are harder to do with solid error analysis than
$\alpha_s$ extractions, but they may be more important in the long run.

\section{Prelattice Quark Mass Results}

\subsection{Quark mass ratios}
The ratios of light quark masses can be investigated with
some degree of reliability
using chiral perturbation theory ($\chi$PT),
which becomes asymptotically exact in the zero quark mass, zero energy
limit of QCD~\cite{Leu90}.
One combination of the light quark mass ratios is 
especially likely to be reliable, since it has been constructed
to have vanishing leading order corrections in $\chi$PT:
\begin{eqnarray}
&&\frac{m_s^2-m_l^2}{m_d^2-m_u^2}=
\frac{M_{K}^2}{M_{\pi}^2}  
	\frac{M_K^2-M_\pi^2}{M_{K^0}^2-M_{K^+}^2+M_{\pi^+}^2-M_{\pi^0}^2}
	\nonumber  \\  
&&\times \left[1+{\cal O}(m_s^2)+{\cal O}\left(e^2\frac{m_s}{m_d-m_u}\right)\right]
\end{eqnarray}

The other combination of quark mass ratios may be obtained at leading
order in chiral perturbation theory from the canonical prediction
of $\chi$PT, that the meson mass squared is proportional to the quark mass:
\begin{equation}\label{msml}
\frac{m_s+m_l}{2m_l}=\frac{M_{K^0}^2}{M_{\pi^0}^2}\left[1+\ldots\right].
\end{equation}
The usefulness of this relation has been called into question by the discovery 
of a symmetry of the chiral Lagrangian which leaves physical predictions 
invariant under simultaneous transformations of the interactions, and of the
quark masses~\cite{kap86}:
\begin{eqnarray}
{\cal L}_\chi &\rightarrow& {\cal L}_\chi'	\\
m_u	&\rightarrow& m_u + \lambda m_s m_d, 	
\end{eqnarray}
 and cyclic in u,d,s.
The effects on the pseudoscalar spectrum of a nonzero $u$ quark mass can be
mimicked exactly by higher order interactions in the $\chi$PT 
Lagrangian and altered $d$ and $s$ quark masses.

In light of this, Leutwyler has assembled a variety of arguments 
to test the size of the corrections to Eq. (\ref{msml})
and in particular the possibility of $m_u=0$ \cite{Leu96}.
He concludes that the corrections are small and
obtains 
\begin{eqnarray}
m_s/m_d  &=& 18.9(8),\\ \label{sd}
m_u/m_d &=& 0.553(43). \label{ud}
\end{eqnarray}

\subsection{Absolute value of the quark masses}

Chiral perturbation theory makes no statement about the 
third combination of light quark masses,
the overall scale.
A large range of results has been obtained with a variety of methods.
See, for example, a compilation of results 
for $m_s$ from Ref. \cite{RPP96} shown in Table \ref{tab:1}.
The most reliable of these prelattice results
 are perhaps those using QCD sum rules,
but even here, the systematic errors of the method are hard to pin down.
It is here that there is the largest spread in existing results and that
lattice methods probably have the most important role to play.

\begin{table}[hbt]
\caption[duh]{Compilation of results for $m_s$ from Ref.~\cite{RPP96}.
If defined,
the renormalization scheme  is $\overline{m}_s(1\ {\rm GeV})$.
}
\begin{tabular}{l|l}
\hline
$m_s$ (MeV)	& method 	\\
\hline
194 (4)		& quark model	\\
118		& sum rules	\\
175 (55)	& sum rules	\\
$>300$		& sum rules	\\
112 (66)	& $\chi$PT + estimate of $\langle \overline{q}q\rangle$	\\
378 (220)	& 
      $\chi$PT + estimate of $\langle \overline{q}q\overline{q}q\rangle$	\\
150		& strange baryon splittings	\\
135		& SU(6)	\\
\hline
\end{tabular}\label{tab:1}
\end{table}

\section{Lattice Quark Mass Determinations}

Lattice determinations of standard model parameters require:
1) fixing the bare lattice parameters from physics, and
2) obtaining the $\overline{MS}$ parameters from these with
short-distance matching calculations.
Spin averaged splittings in the $\psi$ and $\Upsilon$ systems 
are convenient quantities to set the lattice spacing \cite{El-Khadra+al}.
The mesons are small and easy to understand because the
quarks are nonrelativistic.  Light
pseudoscalar meson masses are the most convenient quantities to fix the
light quark masses.	
Chiral symmetry makes them very sensitive to the quark masses,
the mesons are small,
the correlators have good statistics and are easy to fit over long time 
separations.
The correlators  have simple behavior as the quark mass is varied toward zero,
in contrast with unstable vector mesons.
Other quantities must give the same results as the
approximations are removed.

The second piece is the determination of  the
 parameters of the $\overline{MS}$ Lagrangian 
 by matching perturbative, dimensionally regularized
short distance amplitudes to their lattice counterparts.
It is desirable to do the lattice part of such calculations nonperturbatively
as much as possible, to test for the presence of nonperturbative 
short-distance effects and possible poor convergence of perturbation theory.
Such nonperturbative short-distance calculations 
 are harder to design for  quark masses than
for the strong coupling constant.
Such nonperturbative short distance analysis of quark mass extractions
is  currently   less advanced than the analogous investigations 
for the strong coupling constant $\alpha_s$.

The perturbative expression giving the $\overline{MS}$ mass
from the lattice bare mass $m_0$
may be written 
\begin{equation}\label{m0msbar}
\overline{m}(\mu)
= \tilde{m}\left[1+g^2  \left( \gamma_0  \left( \ln \tilde{C}_m-\ln\left( a \mu\right) \right) \right) \right].  \label{eq:star}
\end{equation}
 The mean-field-improved bare mass $\tilde{m}$ is given by
$\tilde{m} = m_0/(1-\frac{1}{12}g^2)$
in perturbation theory, and
$\tilde{m}=m_0/\sqrt[4]{\langle U_P\rangle_{MC}}$
nonperturbatively, if the expectation value of the 
plaquette, 
$\langle U_P\rangle$, 
is used to define 
mean field improvement\cite{LM93}.
$\gamma_0=1/(2\pi^2)$ is the leading quark mass
anomalous dimension.
This coefficients in this expression are well behaved for 
the Wilson action and the ${\cal O}(a)$ improved action
of  Sheikholeslami and  Wohlert (SW) \cite{She85}
(for which $\tilde{C}_m = 1.67$ \cite{Gro84} and  4.72 \cite{Gab91}
respectively).
For staggered fermions,    $\tilde{C}_m$  is 132.9 \cite{Gro84},
leading to renormalization factors of 50--100\%
which are not explained by large tadpole graphs.

\begin{figure}
\epsfxsize=0.45 \textwidth
\epsfbox{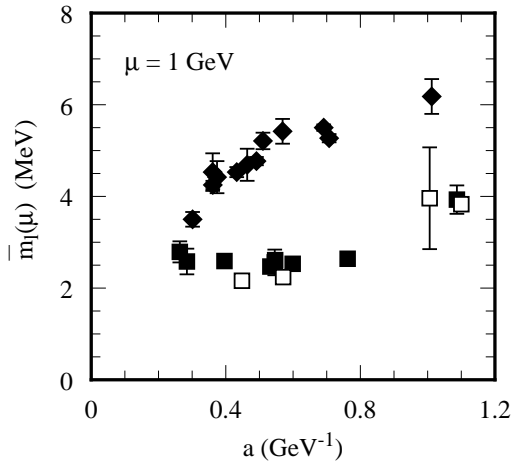}
\caption[old]{
Old lattice quark mass results were reviewed in Ref.~\cite{Uka93}.
Quenched staggered results (dark squares)
show good cutoff independence, but suffer
from huge perturbative corrections.  
Quenched Wilson results (circles) had well-converged perturbation
theory, but large cutoff dependence.
The white squares show some unquenched staggered results.
}
\label{old}
\end{figure}

The status of quark mass extractions as of a few years ago was reviewed in
Refs.~\cite{Uka93,Gup94}.
Some data tabulated in Ref.~\cite{Uka93} are shown in Fig.~\ref{old}.
Unimproved perturbation theory was used, and the lattice spacing was
determined with the $\rho$ mass.
The quenched results for staggered fermions (dark squares)
are relatively independent of the lattice spacing, 
but Eq.~\ref{m0msbar} is untrustworthy  because of the very large 
quantum correction.
The quenched Wilson results (diamonds) 
have much better controlled perturbation theory, but
are much more lattice spacing dependent.
If the results are extrapolated in $a$, they approach the staggered results
 more closely.
However,  the magnitude and origin of  remaining lattice spacing
errors is then unknown.
 Improved actions must be used to investigate this.

\section{Recent Quenched Lattice Results}

\subsection{Unimproved Wilson fermion	 results.}

\begin{figure}
\epsfxsize=.45 \textwidth
\epsfbox{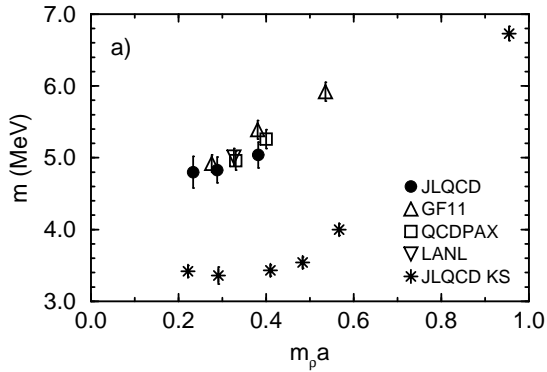}
\caption[jlqcd]{
 Recent unimproved Wilson fermion results for $\overline{m}_l(2\ {\rm GeV})$
from JLQCD compared with previous world data (upper points) \cite{yos96}.
Lattice spacing from $M_\rho$.
JLQCD staggered fermion results are the lower $*$'s.
}
\label{jlqcd}
\end{figure}

New unimproved results have been presented in the JLQCD
contribution to this volume by T. Yoshi\'e \cite{yos96}.
They include a data point at $\beta=6.3$, corresponding to
a smaller lattice spacing
($a=(3.29\ {\rm GeV})^{-1}$) than previous results.
Fig. \ref{jlqcd} shows their Wilson fermion results superimposed on
a subset of the world data.
$M_\rho$  has been used to set the lattice spacing.
 Improved perturbative theory has been used in the renormalizations,
which makes the Wilson results slightly higher than 
in the older analyses, and the staggered results much higher, 
about 50\%.
The new JLQCD Wilson results are statistically consistent with 
previous results,  and so can be combined with them for a consistent
extrapolation of the leading error.
Taken by themselves, however,
they appear to have a somewhat smaller $a$ dependence
then other results.

\begin{figure}
\epsfxsize=0.45 \textwidth
\epsfbox{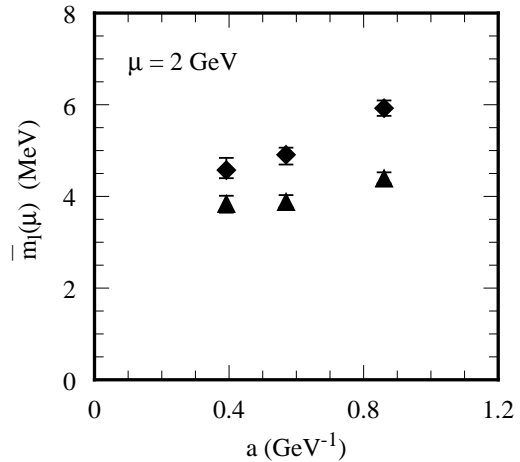}
\caption[fermilab]{
Fermilab light quark masses for Wilson action (circles)
and the tadpole improved SW action (triangles)~\cite{Ono96}.
Lattice spacing  from the 1P--1S splitting in the $\psi$ system.
}
\label{fermilab}
\end{figure}

Wilson data were also presented by Fermilab 
(upper points in Fig.~\ref{fermilab}) \cite{Ono96,gou96}.
These points also lie slightly below and have a smaller slope than
most of the world data in Fig.~\ref{jlqcd}.
 This is due to the fact that 
the Fermilab data use the spin averaged 1P--1S splitting in the $\psi$ system
to determine the lattice spacing,
which is expected to have a small ${\cal O} (a)$ error.
Determining $a$ with the $\rho$ mass is equivalent to calculating
$m_l/m_\rho$.
Since the $ \overline{\psi} \sigma_{\mu\nu} F_{\mu\nu} \psi$
${\cal O} (a)$
correction operator is expected to push spin partners
like the $\pi$ and the $\rho$ apart, 
${\cal O} (a)$ errors in both masses contribute to the $a$ dependence.
When the 1P--1S splitting is substituted for the $\rho$, 
the slope ought to be reduced, as observed.
If $\rho$ mass lattice spacings are substituted into the Fermilab data,
the results line up with the upper points in Fig.~\ref{jlqcd}.

Gupta and Bhattacharya have performed a linear extrapolation
on the world Wilson data and obtain 
$\overline{m}_l(2\ {\rm GeV})= 3.1(3)$ MeV \cite{Gup96}.
They use   slightly different analysis methods from those of Ref.~\cite{yos96},
and obtain a result below that which would be obtained from Fig.~\ref{jlqcd},
which is about 3.8 MeV.
However, the authors say  that  their error estimate
is a ``guess chosen to reflect the various uncertainties discussed \dots
and is not based on systematic analysis''.

JLQCD also presented new results for staggered fermions which left the existing
situation unchanged.
The results are almost lattice independent for $a<(1.5\ {\rm GeV})^{-1}$,
but it is difficult to estimate perturbative uncertainties
 since the perturbative
renormalization factor is so large.  (At $\beta=6.4$, they use
$Z=1.69$, a 69\% correction of which most is unexplained by mean
field improvement.)

\subsection{Improved quenched Wilson fermion results}

To investigate the size of the remaining cut--off
 errors in the Wilson action (${\cal O} (a^2)$, ${\cal O} (a \alpha_s)$,
 ${\cal O} (\alpha_s^2)$, etc.),
it is necessary to remove the leading error with an improved action.
In the SW action, the leading ${\cal O} (a)$ error is removed with 
the addition of the operator
$ \overline{\psi} \sigma_{\mu\nu} F_{\mu\nu} \psi$.
There are several gauge links in the operator, so tadpole improvement
predicts a rather large correction, of order 50\%.

New results with the tadpole improved SW action were presented by
Fermilab \cite{Ono96}.  
The lattice spacing dependence observed in the Wilson points is
substantially reduced, but not completely eliminated.
If this arises from a residual ${\cal O} (a)$ error,
the results extrapolate down to rather near the staggered results.
 (A recent nonperturbative determination of the coefficient of the
SW improvement operator has indicated the possibility of even larger
corrections than the sizable ones 
 indicated by perturbation theory and mean field 
improvement~\cite{Jan96}.)
If it arises from higher order effects in $a$, the answer is close to the 
existing small lattice spacing points, and the discrepancy with staggered
results must be attributed to poorly behaved staggered perturbation theory.

Since a downward trend in $a$ is still present in the tadpole improved
data, we take the improved result at the finest lattice spacing 
as an upper bound in the quenched approximation and a linear extrapolation
as a lower bound.
Adding perturbative uncertainties linearly and other uncertainties in
quadrature gives the quenched result:
\begin{eqnarray}\label{eq:qu}
\overline{m}_l(2\ {\rm GeV})	& = 	& 3.6 (6)\ {\rm MeV}	,\\
\overline{m}_s(2\ {\rm GeV})	& = 	& 95 (16)\ {\rm MeV}.
\end{eqnarray}

Another determination of the strange quark mass with an ${\cal O}(a)$
improved  action has been reported,
in Ref. \cite{All94}.
This determination used a tree-level, rather than a mean-field improved
coefficient for the improvement operator.
They obtained $\overline{m}_s(2\ {\rm GeV})=128 \pm 18$ MeV.
They did not attempt to
 correct for the effects of the remaining lattice  spacing
dependence
or the effects of the quenched approximation.
Much of the discrepancy with the Fermilab
 results arises from fact that we have used
much larger SW improvement
 coefficients, and make an allowance for the fact that 
we continue to find significant cut-off dependence even so.

\section{Corrections to Quark Mass Ratios}

\subsection{Nonlinearities in $m_q$ vs. $M_\pi^2$}

\begin{figure}
%\vskip2.3in
\epsfxsize=0.45 \textwidth
\epsfbox{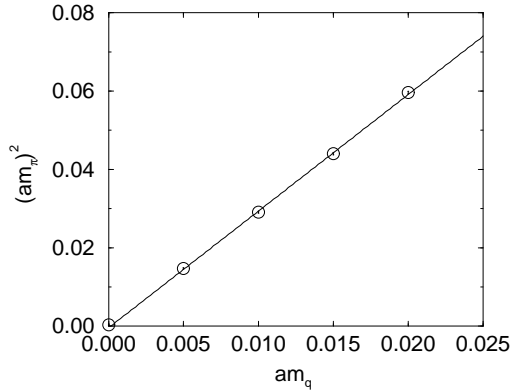}
\caption[nonlin]{
$(aM_\pi)^2$ vs. $am_q$ for staggered fermions at $\beta=6.4$
on a $40^3\times 96$ lattice \cite{jlq96}.
Deviations from linearity are a few \% or less.
}
\label{nonlin}
\end{figure}

There is very little lattice evidence for large deviations from linearity
in the quark masses in Eq. (\ref{msml})
in either quenched or unquenched calculations.
In the most accurate data, the question often seems to be why 
the predictions of chiral symmetry appear linear
at the few~\% level up to such large values of the quark masses.
(See Fig.~\ref{nonlin}.  The largest pion mass is around 850 MeV,
and nonlinearity in $(aM_\pi)^2$ vs. $am_q$ is only a few~\%.)
However,  searching for such deviations is tricky,
especially in the quenched approximation where quenched chiral logarithms
may add spurious nonlinearities at small $m_q$ \cite{Sha96}.
 Existing unquenched calculations have not yet examined carefully the 
case of broken flavor SU(3), $m_s>m_l$, so not all of the higher order
operators of the chiral Lagrangian have yet been tested.

\subsection{Electromagnetic effects} It is in principle simple to determine the contributions of electromagnetism
to hadron masses using numerical simulations with link matrices which
are products of SU(3) and U(1) matrices.
The U(1) phases of electromagnetism are cheap to generate because 
to leading order in $\alpha_{em}$ 
(all that is required for practical purposes),
they can be obtained from  Fourier transforms of Gaussian fluctuations 
in momentum space.
To see the effects of electromagnetism clearly above fluctuations
in the SU(3) field, one wants to use  values of  $\alpha_{em}$
which are larger than the physical value.
It then remains only  to show that electromagnetic effects are still
linear in this region.  
That this indeed holds has been shown recently by Duncan, Eichten,
and Thacker \cite{dun96}.
They have performed a prototype calculation of the $\pi^+ - \pi^0$
splitting at $\beta=5.7$ and with it have obtained $m_u/m_d = 0.51$,
in agreement (so far) with Eq.~\ref{ud}.

\subsection{Can $m_u = 0$?}
The most interesting application of these calculations
is the settling of the question of whether $m_u \equiv 0$ 
 in the real world.
This possibility is fervently desired in spite of all evidence to the
contrary because of its neat solution to the strong CP problem.
Ref.~\cite{kap86} showed that allowing corrections 
as large as 30\% to $\chi$PT
equations such as \ref{msml} made $m_u = 0$ compatible with meson mass data.
(See Fig.~\ref{nonlin}, however.)
The possibility of large instanton-induced flavor mixing effects
has been proposed as a mechanism to generate such corrections in 
QCD~\cite{Geo81}.
Not all the required lattice calculations have been done, but so far,
the lattice evidence is against $m_u=0$.

\section{The Quenched Approximation}\label{quapp}

As the lattice spacing is reduced, while keeping hadronic physics fixed,
the lattice bare couplings evolve according to their anomalous dimensions.
In the quenched approximation, these anomalous dimensions are slightly 
wrong, due to the absence of light quark loops.
The strong coupling constant evolves according to the zero-quark $\beta$ 
function coefficient $\beta_0^{(0)}=11$ rather than the correct three-quark
coefficient $\beta_0^{(3)}=11-2/3 n_f=9$.
Asymptotically, $\alpha_s(\pi/a)$ in the quenched approximation is expected
to be too small by a factor of about 9/11.

Since the short-distance quark mass evolution is 
given by $d\ln m(q)/d \ln(q) = - \gamma_0 \alpha_s/(4\pi)$, 
where $\gamma_0 = 8$,
this implies that the quark mass evolves too slowly
in the quenched approximation, and 
at  small lattice spacings is larger than in real life.
At high energies, running mass evolution is given by
\begin{equation}
\frac{m(q_1)}{m(q_2)} \approx 
	\left( \frac{\alpha_s(q_1)}
		{\alpha_s(q_2)}\right)^{\frac{\gamma_0}{2\beta_0}}.
\end{equation}
To leading log accuracy, therefore, the effect of the absence of quark loops
due to {\em perturbative} effects
on the evolution of the running mass from the strong coupling region
(where $\alpha_s\approx 1$) to the high energy region can be 
approximated by
\cite{Mac94}
\begin{eqnarray}
\frac{m(\pi/a)|_{\rm qu.\ \ \ }}{m(\pi/a)|_{\rm unqu.}}
&\approx& \alpha(\pi/a)^{\frac{\gamma_0}{2}(1/\beta_0^{(0)}-1/\beta_0^{(3)})}\\
&\approx& 1.1 {\ \rm to \ } 1.2,  \label{eq:1.15}
\end{eqnarray}
for $\alpha(\pi/a)\approx 1/4$ to 1/8.

There are also quenching effects arising from the nonperturbative region.
Unlike the case of quarkonium systems, for pions there is no argument that
these should be smaller than the perturbative effects.
The above expression can't be taken as a correction factor,
only as an indication of the direction and order of magnitude of effects to
be expected.
Nonperturbative calculations are required to investigate quenching
effects quantitatively.

Unquenched results for Wilson fermions appear complicated and hard
to interpret.
They differ by as much as a factor of two from quenched results,
and do not seem to reproduce the lattice spacing dependence of the quenched
results as would have been expected (see Ref. \cite{Uka93}).
Since most other unquenched Wilson results, such as those in thermodynamics, 
are also hard to understand compared to staggered results,
I will not attempt to fit unquenched Wilson fermion results into
my general picture.

Some  unquenched staggered results summarized in Ref. \cite{Uka93} 
are
shown in Fig. \ref{old} (white squares) along with
the quenched results.
The unquenched results indeed lie below the quenched results by roughly 
the expected amount, and I will use them
to estimate the effects of  quenching.
The large corrections in staggered fermion mass renormalization
  cancel out in the ratio of the quenched
and unquenched determinations, making this an useful quantity to examine.
To minimize effects due to differences in analysis methods, we estimate
the ratio from the results of a single group, at similar volumes and
lattice spacings (about 0.4 GeV$^{-1}$)  \cite{Ish92,Fuk92}, and obtain
\begin{eqnarray}
\frac{\overline{m}_l(1.0\ {\rm GeV})_{\rm n_f=0}}{\overline{m}_l(1.0\ {\rm GeV})_{\rm n_f=2}}
&\approx& \frac{2.61(9)}{2.16(10)}\\
&=& 1.21(7)  \label{eq:unq}
\end{eqnarray}
Since there are, in fact, three flavors of light quarks in the world 
and not two, I will this ratio as a lower bound on the actual ratio
and use the square (corresponding to four light quarks) as an upper bound.

\section{Synthesis}

Lattice determinations of light quark masses are more difficult than
the analogous determinations of $\alpha_s$.
Pion masses have worse $a$ dependence than the quarkonium splittings.
Finding nonperturbative ways of eliminating the perturbative
relations with the bare lattice quark mass requires more work.

Nevertheless, with a few exceptions, 
most lattice quark mass extractions are consistent with a reasonably
simple picture.
I ignore results with very large $a$ or  very small volume.
Unquenched Wilson results, which are also hard to interpret in most
other quantities,  do not seem to make sense compared with
quenched Wilson results.
Of the remaining results,
the larger magnitude and lattice spacing
 dependence of Wilson results compared with
staggered results is greatly reduced with 
${\cal O} (a)$ improved actions.
The magnitudes are reduced more with tadpole improved SW correction terms
than with tree--level correction terms.
If the remaining discrepancy  arises mainly from 
residual  ${\cal O} (a)$ effects in the improved action,
the true quenched answer lies close to the staggered result,
$\overline{m}_l$ a little over 3 MeV.
If it arises mainly from higher order corrections in the staggered fermion
lattice--$\overline{MS}$ mass conversion (where the leading correction
is 50--100\%), the true answer lies closer to the improved result,
$\overline{m}_l \sim 4$ MeV.

Unquenched staggered results lie somewhat below quenched staggered results,
but by an amount which is reasonable. 
Taking the ratio from Sec.~\ref{quapp}   and using it to make a correction on
our quenched result we obtain
\begin{itemize}
\item $\overline{m}_s(2\ {\rm GeV})$ in the range  54 -- 92  MeV, 
\item $\overline{m}_l(2\ {\rm GeV})$ in the range  2.1 -- 3.5 MeV,
\end{itemize}
for the $\overline{MS}$ masses renormalized at 2 GeV.
The uncertainties most in need of further study are
those associated with lattice spacing dependence and the quenched 
approximation.

\section*{Acknowledgments}

I thank  Brian Gough, Aida El-Khadra, George Hockney,
Andreas Kronfeld, Bart Mertens, Tetsuya Onogi, and Jim Simone
for collaboration on the work of Ref. \cite{Ono96,gou96}.
I thank the Center for Computational Physics in Tsukuba for hospitality
while this paper was written,
and I thank the members of the JLQCD collaboration for useful discussions.
Fermilab is operated by Universities Research
Association, Inc. under contract with the U.S. Department of
Energy.


\begin{thebibliography}{9}

\bibitem{RPP96} Review of Particle Properties, 
    R.M. Barnett et al., Phys. Rev. {\bf D54} (1996) 1.

\bibitem{Leu90} H. Leutwyler Nucl. Phys. {\bf B337} (1990) 108,
and references therein.

\bibitem{kap86} D. B. Kaplan and A. V. Manohar  
   Phys. Rev. Lett. {\bf 56} (1986) 2004.

\bibitem{Leu96} H. Leutwyler,  Phys. Lett. {\bf B378} (1996) 313.

\bibitem{El-Khadra+al} A. X. El-Khadra, G. Hockney, A. S. Kronfeld, and P. B.
	Mackenzie, Phys. Rev. Lett. {\bf 69} (1992) 729.
\bibitem{LM93} G. P. Lepage and P. B. Mackenzie,  Phys. Rev. {\bf
	D48} (1993) 2250.

\bibitem{She85}
  B. Sheikholeslami and R. Wohlert, Nucl. Phys. {\bf B259} (1985) 572.

\bibitem{Gro84} R. Groot, J. Hoek, and J. Smit, Nuc. Phys. {\bf B237} (1984)
	111; and references therein.
\bibitem{Gab91} E. Gabrielli, G. Martinelli, C. Pittori, G. Heatlie,
    C.T. Sachrajda,     Nucl. Phys. {\bf B362} (1991) 475.

\bibitem{Uka93}  A. Ukawa,  Nucl. Phys. B Proc. Suppl.{\bf 30} (1993) 3.
\bibitem{Gup94}  R. Gupta,  Nucl. Phys. B Proc. Suppl.{\bf 42} (1995) 85.

\bibitem{yos96} JLQCD Collaboration (presented by T. Yoshi\'e), 
	these proceedings.

\bibitem{Ono96} T. Onogi, in these proceedings.
\bibitem{gou96}  B. Gough, A. X. El-Khadra, G. Hockney,
	A. S. Kronfeld, P. B. Mackenzie, B. Mertens, T. Onogi, and J. Simone,
	Fermilab preprint  Pub-96/283-T.
\bibitem{Gup96}R. Gupta and T. Bhattacharya, 
     LA-UR-96-1840 (1996)     [HEP-LAT 9605039].

\bibitem{Jan96} K. Jansen, C. Liu, M. L\"uscher, H. Simma, 
    S. Sint, R. Sommer, P. Weisz, and U. Wolff,
    Phys. Lett. {\bf B372} (1996) 275.

\bibitem{All94} C.R. Allton et al., Nucl. Phys. {\bf B431} (1994) 667.
\bibitem{jlq96} The JLQCD collaboration, S. Aoki, private communication.
\bibitem{Sha96} See, for example, the review by S. Sharpe in 
	these proceedings.
\bibitem{dun96}  A. Duncan, E. Eichten, and  H. Thacker, 
    Phys. Rev. Lett. {\bf 76} (1996) 3894.
\bibitem{Geo81} H. Georgi and I. N. McArthur, Harvard University Report
	 HUTP-81/A011 (1981).

\bibitem{Mac94}
    P. B. Mackenzie, Nuc. Phys. B, Proc. Suppl. {\bf 34} (1994) 400.

\bibitem{Ish92}  N. Ishizuka, M. Fukugita, H. Mino, M. Okawa, A. Ukawa,
    Nucl. Phys. Proc. Suppl. {\bf 26} (1992) 284.

\bibitem{Fuk92} M Fukugita et al., Phys. Rev. Lett. {\bf 68} (1992) 761.

\end{thebibliography}
\end{document}